\documentclass{pasj00}

\draft

\begin{document}
\SetRunningHead{S. Kato}{}
\Received{2013/00/00}
\Accepted{2013/00/00}

\title{Resonant Excitation of Tilt Mode in Tidally Deformed Disks} 

\author{Shoji \textsc{Kato}}
\affil{2-2-2 Shikanodai-Nishi, Ikoma-shi, Nara, 630-0114}
\email{kato@gmail.com, kato@kusastro.kyoto-u.ac.jp}

%

\KeyWords{accretion, accrection disks 
          --- dwarf novae
          --- negative superhumps
          --- oscillations
          --- resonance
          --- tilt 
          --- tidal deformation
         } 

\maketitle

\begin{abstract} 
In a previous paper (Kato 2013b), we have shown that in deformed disks a pair of 
trapped oscillation modes can be resonantly excited through couplings with
disk deformation.
In this paper we examine in what cases tilts are excited on
tidally deformed disks by the above-mentioned wave-wave resonant process.
The results show that tilts can be excited in various evolutional stages
of tidally deformed disks, although the wave mode which becomes the pair to  
the tilt and the mode of tidal waves contributing to the resonance change by change 
of disk stages.
\end{abstract}

\section{Introduction}

Lubow (1991) showed that in tidally deformed disks,
there is a mode-mode coupling which leads the disk to eccentric deformation.
This is the so-called 3 : 1 resonance and called the tidal instability.
Lubow (1992) further showed that this 3 : 1 resonance can also excite tilts of disks.

In a similar context, Kato (2004, 2008) examined a wave-wave resonant coupling process
in deformed disks in order to examine the origin of quasi-periodic oscillations (QPOs)
observed in low-mass X-ray binaries.
This wave-wave resonant excitation process in deformed disks was numerically examined by
Ferreira \& Ogilvie (2008) and Oktariani et al. (2010).
Subsequently, Kato et al. (2011) and Kato (2013b) formulated the instability condition 
in a general way and derived the instability conditions in a simple and general form.
Concept of wave energy is of importance to understand the instability.
The instability criterion derived is found to be formally extended to cases of 
magnetized disks (Kato 2013c).

The wave-wave resonant process is essentially the same as Lubow's mode-mode coupling process, and
now found to be an extension of the latter to more general situations, based on a view point of 
wave phenomena.
Really, we can reproduce Lubow's tidal instability, Osaki's precession of one-armed-oscillations,
and Lubow's excitation of tilt at the 3 : 1 resonance also by the wave-wave resonant process
(Kato 2013b).

A brief examination of the wave-wave resonant excitation process 
suggests that the process works in other realistic situations than the 3 : 1 
resonance, especially when excitation of tilt is concerned.
This comes from the facts that in the excitation of tilts 
i) a vertical p-mode oscillation becomes the pair to the tilt mode and 
ii) the local frequency of such oscillation is not just 
integers of the local Keplerian frequency.
From the observational points of view, on the other hand, the presence of negative superhumps has been 
observed in various phases of dwarf novae (e.g., Patterson et al. 1995, Ohshima et al. 2012).
Negative superhumps are usually supposed to be due to tilts of disks, although their origin 
is not understood yet.
[See Osaki and Kato (2013a, b) for extensive analyses of observational data obtained by the Kepler 
telescope, including those of negative superhumps.]

Considering the above situations, we examine in this paper in what situations  
the excitation of tilts is expected by the wave-wave resonant process on tidally deformed disks.
If we restrict our attention only to applications to dwarf navae, examination of tilt excitation
in coplanar systems with circular orbit of the secondary star will be enough.
By considering possible applications to such systems as the Be/X-ray binaries, however, we also examine the cases
where these two planes are misaligned and the orbit is eccentric.
It is noted that what is done in this paper is only for some simplified cases.
To obtain more qualitative results in realistic situations, numerical analyses evaluating
growth rates by calculating coupling terms with detailed informations concerning eigenfunctions of trapped
oscillations  are necessary. 
This is a subject in the near future.

To study the excitation of tilts, we must know what types of tidal waves are present on tidally
deformed disks.
The main factors determining the types of tidal waves are i) the distance between the primary and secondary
stars, ii) eccentricity of the orbit of the secondary star, and iii) the inclination between the disk plane
and the orbital plane.
In appendix 2, the relations among the above factors and the types of tidal waves are summarized. 

It is noted here that in this paper, for simplicity, we have used the coordinate system whose center 
is on the primary star.
If we want to apply the present results quantitatively to real observed systems,
the coordinate system should be changed to the observational one.

\section{Outline of Wave-Wave Resonant Processes in Tidally Deformed Disks}
 
Let us assume that oscillations in disks can be decomposed into normal modes.
The time and angular dependences of the displacement vector, $\mbox{\boldmath $\xi$}(\mbox{\boldmath $r$}, t)$,
associated with a normal mode of the oscillations are factorized as
\begin{equation}
   \mbox{\boldmath $\xi$}(\mbox{\boldmath $r$}, t)=\Re\biggr[\hat{\mbox{\boldmath $\xi$}}(r,z)
                     {\rm exp}(i\omega t)\biggr]
         =\Re\biggr[\breve{\mbox{\boldmath $\xi$}}(r,z)
                     {\rm exp}[i(\omega t-m\varphi)]\biggr],
\label{2.0}
\end{equation}
where $\Re$ denotes the real part.
Here, $\mbox{\boldmath $r$}$ is the cylindrical coordinates ($r, \varphi, z$), whose center is
at the disk center (this is also the center of the primary star) and the $z$-axis is the axis of the 
rotating axis of the disk.
We consider two oscillations specified by $\mbox{\boldmath $\xi$}^{(1)}(\mbox{\boldmath $r$}, t)$ and
$\mbox{\boldmath $\xi$}^{(2)}(\mbox{\boldmath $r$}, t)$.
The set of frequency, azimuthal wavenumber, and vertical node number, i.e., ($\omega$, $m$, $n$), 
of each oscillation is denoted as ($\omega_1$, $m_1$, $n_1$) and ($\omega_2$, $m_2$, $n_2$).
Generally, the radial and vertical components of a normal mode of oscillations 
i.e., $\xi_r(\mbox{\boldmath $r$}, t)$ and $\xi_z(\mbox{\boldmath $r$}, t)$, have different node numbers 
in the vertical direction.
The latter component has one less number compared with the former has. 
Here and hereafter, $n$ is used to denote the node number of $\xi_r$.

Next, let us consider the component of tidal force with azimuthal wavenumber $m_{\rm D}(=\pm 1,
\pm 2,\pm 3)$ and rotational frequency $\omega_{\rm D}$, where $\omega_{\rm D}/\Omega_{\rm orb}=
\pm 1,\pm 2,\pm 3,...$, $\Omega_{\rm orb}$ being the orbital frequency of the secondary 
star around the primary.
Then, the displacement vector, $\mbox{\boldmath $\xi$}^{(\rm D)}(\mbox{\boldmath $r$}, t)$,
associated with the tidal force is characterized by ($\omega_{\rm D}$, $m_{\rm D}$).
Concerning the $z$-dependence of $\mbox{\boldmath $\xi$}^{(\rm D)}(\mbox{\boldmath $r$}, t)$, 
an attention is necessary.
In coplanar systems where the disk plane and the orbital plane of the secondary coincide, 
the tidal potential and thus $\xi^{(\rm D)}_r(\mbox{\boldmath $r$}, t)$ is an even function with respect to $z$, 
while in the case where these two planes
are misaligned, $\xi^{\rm D}_r(\mbox{\boldmath $r$}, t)$ has both even and odd components with respect to $z$.

The two disk oscillations, $\mbox{\boldmath $\xi$}^{(1)}(\mbox{\boldmath $r$}, t)$
and $\mbox{\boldmath $\xi$}^{(2)}(\mbox{\boldmath $r$}, t)$,
can have non-linear resonant couplings through the disk deformation, if the 
following resonant conditions are satisfied\footnote{
In Kato et al. (2011) and Kato (2013a) we have adopted as the resonant conditions
$$
    \omega_2=\omega_1\pm\omega_{\rm D} \quad {\rm and}\quad m_2=m_1\pm m_{\rm D}.    \nonumber
$$
However, since the signs of $\omega$'s and $m$'s can be taken to be negative, the forms of
equations (\ref{2.1}) are convenient for general arguments, and in Kato (2013b) we have adopted 
expressions (\ref{2.1}).
Here and hereafter, we adopt expressions (\ref{2.1}).
}, i.e., 
\begin{equation}
    \omega_1+\omega_2+\omega_{\rm D}=0\quad{\rm and}\quad 
    m_1+m_2+m_{\rm D}=0.
\label{2.1}
\end{equation}
The above resonant conditions are necessary for resonant instability, but not sufficient.
For the resonance to lead instability, additional conditions are necessary.
Of course, the oscillations must have a common propagation region.
Otherwise, the coupling is weak.
More importantly, if ($E/\omega$)'s of the two oscillations have the same signs,
i.e., 
\begin{equation}
       \biggr(\frac{E_1}{\omega_1}\biggr)\biggr(\frac{E_2}{\omega_2}\biggr)> 0,
\label{inst-cond}
\end{equation}
the resonant coupling leads to instability (Kato 2013b).
Here, $E$ is the wave energy defined by, e.g., for the wave of 
$\mbox{\boldmath $\xi$}^{(1)}(\mbox{\boldmath $r$}, t)$,
\begin{eqnarray}
   E_1=\int \frac{1}{2}\omega_1\biggr[\omega_1\rho_0\hat{\mbox{\boldmath $\xi$}}_1^*
            \hat{\mbox{\boldmath $\xi$}}_1
       -i\rho_0\hat{\mbox{\boldmath $\xi$}}_1^*(\mbox{\boldmath $u$}_0\cdot\nabla)
            \hat{\mbox{\boldmath $\xi$}}_1\biggr]dV    \nonumber   \\
   \sim\int \frac{1}{2}\omega_1\biggr[(\omega_1-m_1\Omega)(\hat{\xi}_{1,r}^*\hat{\xi}_{1.r}
        +\hat{\xi}_{1,z}^*\hat{\xi}_{1,z})\biggr] dV,
\label{wave-energy}
\end{eqnarray}
where $\mbox{\boldmath $u$}_0(\mbox{\boldmath $r$})$ is the velocity on the unperturbed disk,
i.e., $\mbox{\boldmath $u$}_0(\mbox{\boldmath $r$})=(0,r\Omega(r),0)$ and the asterisk denotes
the complex conjugate, and
the integration should be performed over the whole volume where the (trapped) oscillations exist.
The condition, $(E_1/\omega_1)(E_2/\omega_2)> 0$, is roughly equal to
$(\omega_1-m_1\Omega)(\omega_2-m_2\Omega)>0$ [see equation (\ref{wave-energy})].
%

To evaluate the growth rate, we must calculate the magnitude of the coupling.
The efficiency of coupling among three modes of oscillations, $\mbox{\boldmath $\xi$}^{(1)}$, 
$\mbox{\boldmath $\xi$}^{(2)}$, and $\mbox{\boldmath $\xi$}^{({\rm D})}$,
is measured by $W$ defined by (Kato 2013b)
\begin{equation}
     W=\int \hat{\mbox{\boldmath $\xi$}}^{(1)}\cdot\mbox{\boldmath $C$}(\hat{\mbox{\boldmath $\xi$}}^{(2)},
          \hat{\mbox{\boldmath $\xi$}}^{({\rm D})})dV
      =\int \hat{\mbox{\boldmath $\xi$}}^{(2)}\cdot\mbox{\boldmath $C$}(\hat{\mbox{\boldmath $\xi$}}^{(1)},
          \hat{\mbox{\boldmath $\xi$}}^{({\rm D})})dV,
\label{2.1a}
\end{equation}
where  
$\mbox{\boldmath $C$}(\hat{\mbox{\boldmath $\xi$}}^{(2)},\hat{\mbox{\boldmath $\xi$}}^{({\rm D})})$, 
for example,
is the nonlinear terms of the wave equation of $\hat{\mbox{\boldmath $\xi$}}^{(1)}$, representing the 
effects  of coupling between $\hat{\mbox{\boldmath $\xi$}}^{(2)}$ 
and $\hat{\mbox{\boldmath $\xi$}}^{({\rm D})}$ on $\hat{\mbox{\boldmath $\xi$}}^{(1)}$. 
The volume integration of the product $\hat{\mbox{\boldmath $\xi$}}^{(1)}$ and 
$\mbox{\boldmath $C$}(\hat{\mbox{\boldmath $\xi$}}^{(2)},\hat{\mbox{\boldmath $\xi$}}^{({\rm D})})$ is 
related to growth of the oscillation of $\mbox{\boldmath $\xi$}^{(1)}$ (kato 2013b).

An explicit expression for $W$ is (Kato 2004, 2008, 2013b)
\begin{equation}
    W= W^{\rm \psi}+W^{\rm P}+W^{\rm T},
\label{2.2}
\end{equation}
where $W^{\rm \psi}$, $W^{\rm P}$, and $W^{\rm T}$ are expressed, respectively, 
in the case of isothermal oscillations, as
\begin{equation}
    W^{\rm \psi}=-\int \rho_0\xi_i^{(1)}\xi^{(2)}_j\xi^{\rm D}_k
        \frac{\partial^3\psi_0}{\partial r_i\partial r_j\partial r_k}dV,
\label{2.3}
\end{equation}
\begin{equation}
    W^{\rm P}=\int p_0\frac{\partial\xi^{(1)}_i}{\partial r_k}
       \biggr[\frac{\partial\xi^{(2)}_k}{\partial r_j}\frac{\partial\xi^{({\rm D})}_j}{\partial r_i}
        +\frac{\partial\xi_j^{(2)}}{\partial r_i}\frac{\partial\xi^{({\rm D})}_k}{\partial r_j}
        \biggr] dV,
\label{2.4}
\end{equation}
\begin{equation}
      W^{\rm T}=-\int \rho_0\xi^{(1)}_i\xi^{(2)}_j\frac{\partial^2\psi_{\rm D}}{\partial r_i\partial r_j}
             dV,
\label{2.5}
\end{equation}
where $\rho_0(\mbox{\boldmath $r$})$, $p_0(\mbox{\boldmath $r$})$, and $\psi_0(\mbox{\boldmath $r$})$ 
are, respectively, density, pressure, and
gravitational potential in the unperturbed disk,
and $\psi_{\rm D}$ is the part of gravitational potential perturbation by 
the tidal force.
It is important to note that the roles of $\hat{\mbox{\boldmath $\xi$}}^{(1)}$ and 
$\hat{\mbox{\boldmath $\xi$}}^{(2)}$ in $W$ are commutative, as shown in equation (\ref{2.1a}),
which can be shown directly from equations (\ref{2.3}) -- (\ref{2.5}).
This commutability is the reason why the instability criterion is written as a simple form as $(E_1/\omega_1)
(E_2/\omega_2)>0$.
The coupling term, $W$, has no influence on the stability criterion: it affects only growth rate.
The growth rate of the $\omega_1$- and $\omega_2$-oscillations is given by (Kato 2013)
\begin{equation}
     \biggr(\frac{\omega_1\omega_2}{16E_1E_2}\biggr)^{1/2}\vert\Im (W)\vert,
\label{growth}
\end{equation}
where $\Im(W)$ denotes the imaginary part of $W$.
It is noted that the growth rate is proportional to the amplitude (a complex quantity in general) of the tidal deformation.

Finally, it is noted that in the case of coplanar systems, the difference between $n_1$ and $n_2$ 
needs to be zero or even for oscillations to resonantly interact.
This is because if the difference is odd, the products of $\mbox{\boldmath $\xi$}^{(1)}$ and 
$\mbox{\boldmath $\xi$}^{(2)}$ in $W$ is an odd function with respect to $z$, and $W$ vanishes since
$\mbox{\boldmath $\xi$}^{({\rm D})}$ is an even function in coplanar systems.
This can be shown by detailed examinations of the expression for $W$. 
In the case of misaligned systems, however, $\mbox{\boldmath $\xi$}^{({\rm D})}$ has both even and odd
components with respect to $z$, and thus $W$ does not vanish generally even if the difference of 
$n_1$ and $n_2$ is odd.

\section{Tilt Mode Trapped}

Here and hereafter, the $\omega_1$-oscillation is taken to be the tilt mode [in other terminology, 
it is the mode of the corrugation wave (e.g., Kato 2001, Kato et al. 2008)].
That is, we adopt $m_1=1$ and $n_1=1$.
The local dispersion relation\footnote{
The local dispersion relation for isothermal perturbations in vertically isothermal
disks is (e.g., Okazaki et al. 1987, Kato 2001, Kato et al. 2008)
$$
   [(\omega-m\Omega)^2-\kappa^2][(\omega-m\Omega)^2-n\Omega_\bot^2]=
       c_{\rm s}^2k^2(\omega-m\Omega)^2,
$$
where $\kappa$ and $\Omega_\bot$ are, respectively,  the horizontal and vertical epicyclic frequencies,
and $k$ is the radial wavenumber of the oscillations.
This local dispersion relation shows the presence of oscillation modes propagating in the
region of $(\omega-m\Omega)^2>n\Omega_\bot^2 (n=1,2,3...)$, since $\Omega_\bot$ is 
always larger than $\kappa$.
}
shows that this mode has a propagation region 
in the radial region specified by $\omega_1-\Omega<-\Omega_\bot$, where $\Omega_\bot(r)$ is
the vertical epicyclic frequency and larger than the angular velocity of disk rotation, $\Omega(r)$,
in tidally deformed disks.
Since $\Omega-\Omega_\bot< 0$ and tends to zero in the innermost region of the disks, 
the tilt mode with a given frequency $\omega_1(<0)$ is trapped inside a radius, say $r_{\rm L1}$, 
where $\omega_1$ becomes equal to $\Omega-\Omega_\bot$, 
i.e., $\omega_1=(\Omega-\Omega_\bot)_{\rm L1}$, the subscript L1 denoting the value at $r_{\rm L1}$.   
This is shematically shown in figure 1.\footnote{
If $\omega_1$ is smaller than the value of $\Omega-\Omega_\bot$ at the disk outer edge, 
the propagation region is terminated at the the disk edge.
}
It should be noted that this $\omega_1$-oscillation has negative value of $E_1/\omega_1$,
i.e., $E_1/\omega_1<0$, since $\omega-\Omega<0$ [see equation (\ref{wave-energy})]. 

\begin{figure}
\begin{center}
    \FigureFile(80mm,80mm){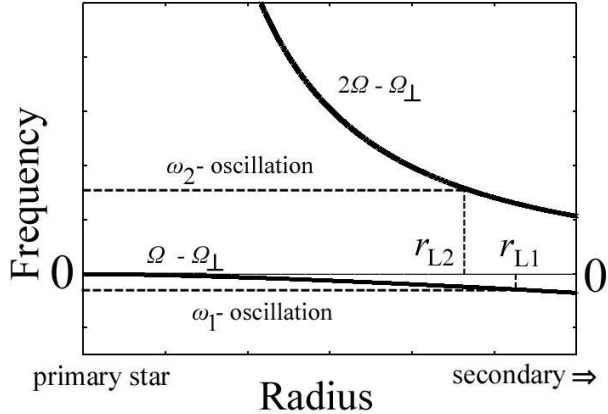}
\end{center}
\caption{Schematic diagram showing propagation regions of $\omega_1$-oscillation (tilt mode)
and $\omega_2$-oscillation.
The scales of coordinates are arbitrary, and are not linear.
The set ($m_1$, $n_1$) of the $\omega_1$-oscillation is (1, 1) and that adopted for 
$\omega_2$-oscillation is (2,1)
The outer edges of the propagation regions of $\omega_1$- and $\omega_2$-oscillations are denoted by
$r_{\rm L1}$ and $r_{\rm L2}$, respectively.
}
\end{figure}

Here, we derive a rough relation between $\omega_1$ (or $r_{\rm L1}$) and temperature 
(or acoustic speed $c_{\rm s}$) in the disks by the WKB method.
(More careful derivation will be made in a subsequent paper.)
The outline of the method is give in Appendix 1.
The results show that the trapping condition (capture condition) is given by
\begin{equation}
     \int_{r_{\rm i}}^{r_{\rm L1}} Q^{1/2}dr =\pi\biggr(n_{\rm r}+\frac{3}{4}\biggr) \qquad (n_{\rm r}=0,1,2,...),
\label{2.6}
\end{equation}
where
\begin{equation}
     Q=\frac{[(\omega-m\Omega)^2-\kappa^2][(\omega-m\Omega)^2-n\Omega_\bot^2]}
            {c_{\rm s}^2(\omega-m\Omega)^2}.
\label{2.7}
\end{equation}
Hereafter we restrict our attention to the fundamental oscillation mode ($n_{\rm r}=0$) 
in the radial direction.
The outer edge of the propagation region, $r_{\rm L1}$, is the
turning point of $Q$ where $Q(r)$ changed its value from positive to negative.
In the trapping condition (\ref{2.6}) 
the radial velocity associated with the oscillation, $u_r$,  has been taken to 
vanish at the inner edge.\footnote{
If the radial gradient of the radial velocity associated with the oscillation is taken to vanish
at ther inner edge, i.e., $du_r/dr=0$, the right-hand side of equation (\ref{2.6}) is
$\pi(n_{\rm r}+1/4)$ instead of $\pi(n_{\rm r}+3/4)$.
}
In the present problem, $m=1$ and $n=1$, since we are interested in the tilt mode,
and we adopt\footnote{
Equations (\ref{2.8}) are crude approximations, based on an expansion of the tidal potential in $r/a$,
as pointed out by the referee.
More accurate expressions are necessary for more detailed calculations.
}  
\begin{equation}
        \Omega-\kappa=\frac{3}{4}q\Omega\biggr(\frac{r}{a}\biggr)^3,  \quad
        \Omega-\Omega_\bot=-\frac{3}{4}q\Omega\biggr(\frac{r}{a}\biggr)^3
\label{2.8}
\end{equation}
where $q$ is the mass ratio of the secondary to primary stars, and $a$ is the mean 
separation between the primary and secondary stars. 
It is noted that since the outer edge of the trapping region, $r_{\rm L1}$, is the turning
point of $Q(r)$,
$r_{\rm L1}$ and $\omega_1$ are related by $Q(r_{\rm L1})=0$.
In the present problem, the relation is 
\begin{equation}
         \omega_1=(\Omega-\Omega_\bot)_{\rm L1}.
\label{trapping-relation}
\end{equation}
This relation between trapping radius (capture radius), $r_{\rm L1}$, and eigenfrequency, $\omega_1$, is shown 
in figure 2 for three cases of $q=0.1$, 0.2, and 0.3.

For simplicity, the acoustic speed $c_{\rm s}$ in disks is taken to be constant,
and the inner edge $r_{\rm i}$ is taken to be zero.
Then, the trapping condition [equation (\ref{2.6})] is a relation between $r_{\rm L1}/a$ 
and $c_{\rm s}/(GM/a)^{1/2}$
with a parameter $q$, where $G$ is the gravitational constant and $M$ is the mass of the primary star.
The relation is shown in figure 3 for three cases of  $q=0.1$, 0.2, and 0.3.

\begin{figure}
\begin{center}
    \FigureFile(80mm,80mm){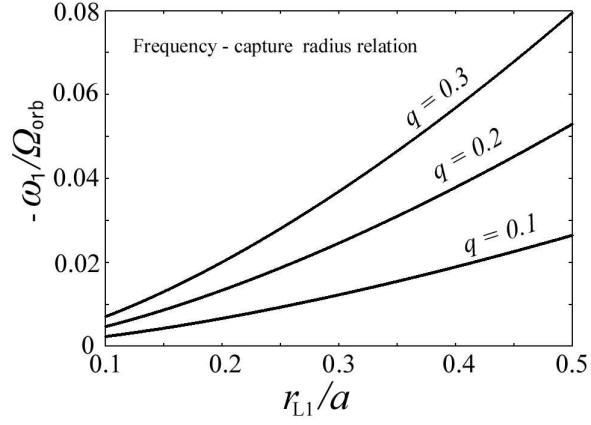}
\end{center}
\caption{
The relation between the frequency, $\omega_1/\Omega_{\rm orb}$, and the capture (trapping) radius, 
$r_{\rm L1}$, of the tilt mode. 
Three cases of the mass ratio between the secondary to the primary, i.e., $q=0.1$, 0.2, and
0.3, are shown. 
}
\end{figure}
\begin{figure}
\begin{center}
    \FigureFile(80mm,80mm){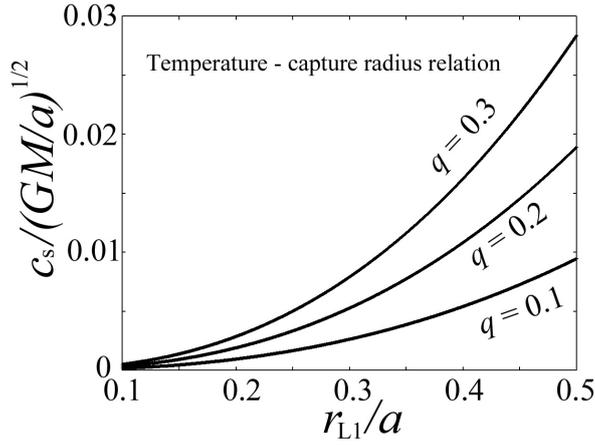}
\end{center}
\caption{Capture radius (trapping radius) $r_{\rm L1}$ -- acoustic speed $c_{\rm s}$ relation 
obtained by the WKB method for the tilt mode with $n_{\rm r}=0$.
Three cases of $q=$ 0.1, 0.2, and 0.3 are shown.
}
\end{figure}
 
\section{Counterpart of Tilt Mode, i.e., $\omega_2$-Oscillation}
 
For the $\omega_1$-oscillation (tilt) mentioned above to be excited by the
wave-wave resonant process, the $\omega_2$-oscillation must also have a negative
value of $E_2/\omega_2$ so that the excitation condition 
$(E_1/\omega_1)(E_2/\omega_2)>0$ is satisfied.
Here and hereafter, we consider vertical p-mode oscillations propagating 
their inner propagation region, since these oscillations have $E_2/\omega_2<0$ there.
In the case of isothermal oscillations in vertically isothermal disks,
their propagation region is specified by $\omega_2<m_2\Omega-\sqrt{n_2}\Omega_\bot$
(see the local dispersion relation given in footnote 2).

Hereafter, we concentrate our attention on the $\omega_2$-oscillations with $n_2=1$, or $n_2=2$,
or $n_2=3$. 
If $m_2\geq 2$, $\omega_2$ is positive, and the propagation region of these oscillations is between 
the inner edge of the disk and the radius where $\omega_2=m_2\Omega-\sqrt{n_2}\Omega_\bot$.
The latter radius is denoted by $r_{\rm L2}$, i.e., 
$\omega_2=(m_2\Omega-\sqrt{n_2}\Omega_\bot)_{\rm L2}$.
The propagation region of the $\omega_2$-oscillation with $m_2=2$ and $n_2=1$ is schematically shown in figure 1. 

Different from the case of $\omega_1$-oscillations, we do not discuss here the trapping
condition of $\omega_2$-oscillations by the following reasons.
The frequency of the $\omega_2$-oscillation which will be considered here is much higher
than that of the $\omega_1$-oscillation, and is comparable with the frequency of disk rotation.
Related to this, the trapped oscillations will be higher overtones in the radial
direction, i.e., they will have many nodes in the radial direction.
In other words, for any frequency $\omega_2$ required from the resonant condition, 
there will always be a trapped oscillation with frequency close to $\omega_2$.
Hence, we assume here that the $\omega_2$-oscillation required from the resonant condition 
roughly satisfies the trapping conditions. 
 
\section{Reconant Conditions and Trapping (Capture) Radius}

Next, let us consider the resonant conditions.
The frequency $\omega_1$ and the outer trapping radius (capture radius) of the $\omega_1$-oscillation, 
$r_{\rm L1}$, are related by $\omega_1=(\Omega-\Omega_\bot)_{L1}$, and also $\omega_2$ and $r_{\rm L2}$
are related by $\omega_2=(m_2\Omega-\sqrt{n_2}\Omega_\bot)_{\rm L2}$, as mentioned before.
Furthermore, we consider the tidal wave whose frequency $\omega_{\rm D}$ is $n\Omega_{\rm orb}$ 
($n=\pm 1,\pm 2,\pm 3,...$) and
whose azimuthal wavenumber is $m_{\rm D}$.
Then, the resonant condition, $\omega_1+\omega_2+\omega_{\rm D}=0$, is written as
\begin{equation}
      (\Omega-\Omega_\bot)_{\rm L1}+(m_2\Omega-\sqrt{n_2}\Omega_\bot)_{\rm L2}
           +n\Omega_{\rm orb}=0,
\label{resonance1}
\end{equation}
and
$m_1+m_2+m_{\rm D}=0$ gives
\begin{equation}
            m_{\rm D}=-(1+m_2).
\label{resonance2}
\end{equation}

The next problem is to know in what case of the $r_{\rm L1}$ and $r_{\rm L2}$ relation the 
growth rate of resonant oscillations become maximum.
The growth rate is given by equation (\ref{growth}) and can be calculated by using the coupling constant $W$,
with eigen-functions of $\omega_1$- and $\omega_2$-oscillations and the form of tidal wave.
This will be done in the near future.
The following considerations concerning the limiting case of zero-temperature disks, however,
suggest that the growth rate will be high when $r_{\rm L1}$ and $r_{\rm L2}$ are close, i.e.,
$r_{\rm L1}\sim r_{\rm L2}$.

Let us first consider the non-linear coupling between the $\omega_1$-oscillation (with $m_1=1$ and $n_1=1$) 
and the tidal wave with $\omega_{\rm D}$ (with an arbitray $m_{\rm D}$ and $n_{\rm D}=0$).
The coupling gives rise to a vertical p-mode oscillation of frequency $\omega_1+\omega_{\rm D}$
with azimuthal wavenumber $m_1+m_{\rm D}$, the vertical node number being unity.
In the limit of zero-temperature disks, the disks respond resonantly 
to this vertical forced oscillation at
the radii where $[(\omega_1+\omega_{\rm D})-(m_1+m_{\rm D})\Omega]^2-\Omega_\bot^2=0$ is satisfied
(e.g., Kato 2008).
If we introduce $\omega_2$ and $m_2$ defined by the resonant conditions, $\omega_1+\omega_2+\omega_{\rm D}=0$
and $m_1+m_2+m_{\rm D}=0$, 
the radii of the above resonant response are specified by $(\omega_2-m_2\Omega)^2-\Omega_\bot^2=0$.
One of two radii which satisfy this condition is $r_{\rm L2}$.
Similarly, let us consider the coupling between the $\omega_2$-oscillation and the tidal wave.
This gives rise to the $\omega_1$-oscillation with $m_1=1$ and $n_1=1$.
To this forced oscillation with frequency $\omega_1$, the disk sharply responds at the radii defined by
$(\omega_1-\Omega)^2-\Omega_\bot^2=0$ ($m_1=1$ and $n_1=1$).
One of the radii is $r_{\rm L1}$.
These considerations
show that in the limiting case of zero-temperature disks, the resonant coupling between 
$\omega_1$- and $\omega_2$-oscillations through the disk deformation occurs if $r_{\rm L1}=r_{\rm L2}$ can be realized
(Lubow 1991,1992).
In disks with finite temperature, however, the resonance is not sharply restricted only to the case of
$r_{\rm L1}=r_{\rm L2}$, since the resonant region is widened by temperature effects (e.g.,
Meyer-Vernet \& Sicardy 1987).
The growth rate, however, will be high in the case where $r_{\rm L1}\sim r_{\rm L2}$.

Taking $r_{L1}\sim r_{L2}$ and representing both radii by a symbol $r_{\rm L}$, 
we have from equation (\ref{resonance1})
\begin{equation}
     \frac{\Omega_{\rm L}}{\Omega_{\rm orb}}\sim -\frac{n}{m_2-\sqrt{n_2}},
\label{resonance3}
\end{equation}
if the difference between $\Omega$ and $\Omega_\bot$ is neglected, where $\Omega_{\rm L}$ is the value of 
$\Omega$ at $r=r_{\rm L}$.
It is noted that $n$ needs to be a negative integer in the present case, since we are interested here in cases
where $m_2-\sqrt{n_2}$ is positive.
In the case of dwarf navae, the disk is known to be roughly truncated by the tidal instability of 
the so-called 3 : 1 resonance at the radius where $\Omega/\Omega_{\rm orb}\sim 3 : 1$ (Lubow 1991).
Hence, hereafter, we restrict our attention only to resonances which occur inside the radius of
$\Omega_{\rm L}/\Omega_{\rm orb}=3$, i.e., to the cases where $\Omega_{\rm L}/\Omega_{\rm orb}>3.0$.
Furthermore, resonances which occur only in misaligned systems are considered separately from those that
occur both in coplanar and misaligned systems.

\begin{table}
  \caption{Normalized capture radius (trapped radius), $r_{\rm L}/a$ of tilts
  for various set of parameters of the $\omega_2$- and tidal oscillations}
  \begin{center}
    \begin{tabular}{cccccccc}
      \hline
    inclination &  ($m_2$, $n_2$) & $n(\equiv \omega_{\rm D}/\Omega_{\rm orb})$ 
             & $m_{\rm D}$ &  $\Omega_{L}/\Omega_{\rm orb}$  & $r_{\rm L}/a$ & $R/D$ & eccentricity \\
        \hline\hline
    $\delta=0$ &  (2,1) &  -3 &  -3 &  3  &    0.48   & $P_3$  & $e^0$ \\
               &        &  -4 &  -3 &  4  &    0.40   & $P_3$  & $e^1$ \\
               &        &  -5 &  -3 &  5  &    0.34   & $P_3$  & $e^2$ \\
  \hline
               &  (2,3) &  -1  & -3 & 3.73 &   0.42   & $P_3$  & $e^2$ \\
               &        &  -2  & -3 & 7.46 &   0.26   & $P_3$  & $e^1$ \\
               &        &  -3  & -3 & 11.19 &  0.20   & $P_3$  & $e^0$ \\
  \hline\hline
     $\delta\not= 0$ &  (2,2)  &  -2  & -3  &  3.41 &  0.44    & $P_4$ & $e^0$ \\
                     &         &  -3  & -3  & 5.12  &  0.34    & $P_4$ & $e^1$ \\
                     &         &  -4  & -3  & 6.83  &  0.28    & $P_4$ & $e^0$ \\                   
    \end{tabular}
  \end{center}
\end{table}

\subsection{Coplanar Cases}

Two cases of $(m_2, n_2)=(2,1)$ and (2,3) are considered.
In the former case we have $\Omega_{\rm L}/\Omega_{\rm orb}\sim -n$ [see equation (\ref{resonance3})], 
which is equal to or larger than 3 for $n\leq -3$.
For three cases of $n=-3$, $-4$, and $-5$, 
some characteristic quantities related to the resonant radius ($r_{\rm L}/a$)
and the mode of tidal wave required ($m_{\rm D}$ and $\omega_{\rm D}$) are shown 
in the upper part (from the second to fourth lines) of table 1.
For the resonance shown in the second line ($n=-3$), the tidal waves required are those with 
$\omega_{\rm D}=-3\Omega_{\rm orb}$ and $m_{\rm D}=-3$.
These tidal waves are practically equivalent to those with $\omega_{\rm D}=3\Omega_{\rm orb}$
and $m_{\rm D}=3$, since there is no change of waves by changing the signs of $\omega_{\rm D}$ and
$m_{\rm D}$ simultaneously.
If the orbit of the secondary star around the primary is circular, the tidal waves which remain 
in the limit of the secondary being far away from the primary is two-armed ($m_{\rm D}=2$)
with $\omega_{\rm D}=2\Omega_{\rm orb}$ [see the terms of $P_2$ in the limit of coplanar systems 
($\delta=0$) in equation (\ref{P_2approx})
in appendix 2].
If the secondary is closer to the primary, three-armed ($m_{\rm D}=3$) tidal waves with 
$\omega_{\rm D}=3\Omega_{\rm orb}$ appear [see the terms of $P_3$ of equation (\ref{P_3approx})
in appendix 2].
The tidal waves required in the case of the second line in table 1 corresponds to this case,
and thus symbols $P_3$ and $e^0$ (no eccentricity) are indicated in the last two columns 
on the second line of table 1.
It is noted that the tilt by this resonance is nothing but what considered by
Lubow (1992) and briefly mentioned by Kato (2013b).

The resonance of the third line, i.e., $n=-4$ and $m_{\rm D}=-3$, shows that $\omega_{\rm D}$ required 
for the resonance is $\omega_{\rm D}=-4\Omega_{\rm orb}$ [see equation (\ref{resonance3})], and $m_{\rm D}$ 
required is $-3$.
That is, $\vert\omega_{\rm D}/\Omega_{\rm orb}\vert$ required is larger than $\vert m_{\rm D}\vert$. 
Deviation of $\omega_{\rm D}$ from $\omega_{\rm D}/\Omega_{\rm orb}=m_{\rm D}$ cannot be realized
as long as $e=0$, i.e., 
eccentric orbits of the secondary are necessary.
In the case of eccentric orbits, the deviation of the tidal force from a form proportional to
${\rm exp}[i(m_{\rm D}\Omega_{\rm orb}t-m_{\rm D}\varphi)]$ is realized by two effects.
One is that the motion of the secondary on the orbit is not at a constant pace
[see equation (\ref{tildphi}) in appendix].
The second is that the distance between the primary and the secondary, $D$, changes with time
[see equation (\ref{aoverD-2}) in appendix].
If these are taken into account till the terms of $e^1$, the tidal force $\psi_{\rm D}$, 
which is given by equation (\ref{psiD3}) in appendix, has terms proportional to
${\rm exp}[i(m_{\rm D}\pm 1)\Omega_{\rm orb}t-im_{\rm D}\varphi]$.
That is, the tidal waves with $\omega_{\rm D}=(m_{\rm D}\pm 1)\Omega_{\rm orb}$ appear in addition
to those with $\omega_{\rm D}=m_{\rm D}\Omega_{\rm orb}$.
The case shown in the third line of table 1 requires such effects of eccentricity $e$, and thus $e^1$ 
is indicated in the last column of table 1.
If the expansion with respect to $e$ is taken into account till the terms of $e^2$, 
tidal waves with $\omega_{\rm D}=(m_{\rm D}\pm 2)\Omega_{\rm orb}$ appear in addition to
those with $\omega_{\rm D}=m_{\rm D}\Omega_{\rm orb}$ 
and $\omega_{\rm D}=(m_{\rm D}\pm  1)\Omega_{\rm orb}$.
The fourth line in table 1 is a case where such tidal  waves are required,
and $e^2$ is indicated in the last column of table 1.   

In the case of $(m_2, n_2)=(2,3)$, we have $\Omega_{\rm L}/\Omega_{\rm orb}\sim -n/(2-\sqrt{3})$
[see equation (\ref{resonance3})],
and the ratio $\Omega_{\rm L}/\Omega_{\rm orb}$ becomes larger than 3.0 for $n\leq -1$.
Some characteristic quantities related to the resonant radius and the mode of tidal wave required 
are shown in the middle part (the fifth to seventh lines) of table 1.

The radial positions of resonant excitation of the tilt mode by various processes
mentioned above are plotted on the $r_{\rm L}$ -- $c_{\rm s}$ diagram in figure 4.
It is noted that the plots are on the $r_{\rm L}$ -- $c_{\rm s}$ curve given in figure 3.

\begin{figure}
\begin{center}
    \FigureFile(120mm,120mm){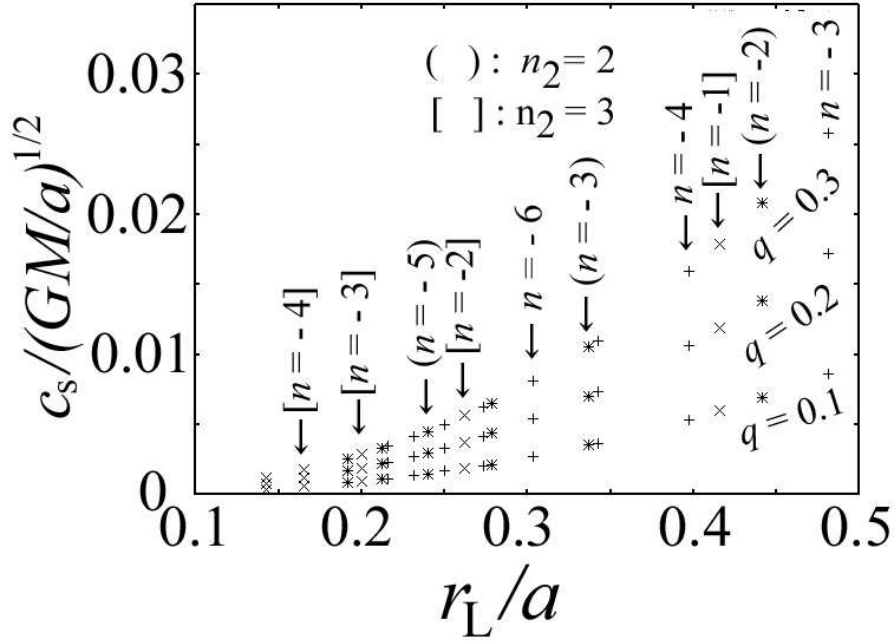}
\end{center}
\caption{Positions of various resonance on the radius - temperature diagram.
Two capture radii of the set of two oscillations, $r_{\rm L1}$ and $r_{\rm L2}$, are taken to be the 
same and denoted by $r_{\rm L}$.
The azimuthal wavenumber of the $\omega_2$-oscillation is taken to be $m_2=2$ in all cases,
and for $n_2$ of the $\omega_2$-oscillation three cases are considered.
The resonant positions in the cases of $n_2=1$ shown by symbol "+", 
and those in the cases of $n_2=2$ by symbol "$\times$", and those in the cases of $n_2=3$ by "$\Large *$".
The value of $n(\equiv \omega_{\rm D}/\Omega_{\rm orb})$ required is indicated  for each resonance by
attaching symbols after the arrow.
The three sequences of the plots show the difference of the mass ratio of the two stars,
i.e., $q=0.3$, 0.2, and 0.1 from the upper sequence to the lower one.
It is noticed that the resonant points are along the $r_{\rm L}$ -- $c_{\rm s}$ curves given in figure 3.
}
\end{figure}

\subsection{Misaligned Cases}

If the disk and orbital planes are misaligned (i.e., $\delta\not= 0$), the tilt mode ($m_1=1$ and $n_1=1$)
can be resonantly excited even when the $\omega_2$-oscillation to be paired with the tilt is plane-symmetric
(i.e., $n_2=$ even), since the tidal force has a plane-asymmetric term 
[see that $P_2$, $P_3$, ... given in appendix 2 have terms proportional to ${\rm sin}\ \gamma$, which are
those of odd powers of $z$].
Here, we consider, in particular, the case of $(m_2, n_2)=(2,2)$.
In this case we have $\Omega_{\rm L}/\Omega_{\rm orb}=-n/(2-\sqrt{2})$ [see equation (\ref{resonance3})].
Three cases of $n=-2$, $-3$, and $-4$ are considered and some characteristic quantities related to
the resonant radius and the mode of tidal wave
required are shown in the lower part (the eighth to tenth lines) of table 1.\footnote{
It is noted that the resonant excitation discussed in the coplanar cases are still
present in misaligned systems.
The resonant processes to be mentioned below are additions to them.
}

Let us first consider the case where the orbit of the secondary is circular ($e=0$).
As is shown in appendix 2, the tidal force resulting from $\delta\not= 0$ (but $\delta\ll 1$ is assumed)
has a term proportional to
${\rm exp}[i(2\Omega_{\rm orb}t-\varphi)]$ in $P_2$ [see equation (\ref{P_2approx})], terms proportional to
${\rm exp}[i(3\Omega_{\rm orb}t-2\varphi)]$ and ${\rm exp}[i(\Omega_{\rm orb}t-2\varphi)]$ in $P_3$
[see equation (\ref{P_3approx})], and terms proportional to 
${\rm exp}[i(4\Omega_{\rm orb}t-3\varphi)]$ and ${\rm exp}[i(2\Omega_{\rm orb}t-3\varphi)]$ in $P_4$, although
a detailed expression for $P_4$ is not presented in appendix 2.

The tidal force required in the case of eighth line of table 1 is proprotional to 
${\rm exp}[i(2\Omega_{\rm orb}t-3\varphi)]$ (i.e., $n= -2$ and $m_{\rm D}=-3$).
As mentioned in the above paragraph, the tidal force resulting from $\delta\not= 0$ has such a term
in $P_4$ with $e=0$.
That is, the tidal force required in the case of eighth line of table 1 appears in $P_4$ with $e=0$.
Thus, $P_4$ and $e^0$ are indicated in the last two colums of the eighth line of table 1.
In the case of nineth line of table 1, however, eccentricity is necessary, since $n$ and $m_{\rm D}$
required in this case are the same (i.e., $n=-3$ and $m_{\rm D}=-3$, in other words, $n=3$ and $m_{\rm D}=3$),
and such a tidal force does not appear from the term resulting from $\delta\not=0$ if $e=0$.
Such a tidal force, however, appears in $P_4$ if the effects of eccentricity is taken into
account till the terms proportional to $e^1$.
To emphasize this point, $P_4$ and $e^1$ are noticed in the last two colums of the nineth line of table 1.
The tidal force required in the case of the tenth line is proportional to ${\rm exp}[i(4\Omega_{\rm orb}t-3\varphi)]$.
Such force is present in $P_4$ even when $e=0$, as mentioned in the above paragraph.
Thus, $P_4$ and $e^0$ are noticed in the last two colums of the tenth line of table 1.

The radial positions of resonant excitation of tilt in the cases 
mentioned above are also plotted on the $r_{\rm L}$ -- $c_{\rm s}$ diagram in figure 4.

Although tilts which are excited at various capture radius, $r_{\rm L}$, are listed in table 1 and figure 4,
this does not mean that all of them are excited to an observable level in dwarf novae.
In close binary systems such as dwarf novae, the eccentricity of the orbits of the secondary star is small,
i.e., $e\sim 0$.
Hence, in the cases labelled by $e^1$ or $e^2$ in the last column of table 1, the amplitude of the tidal deformation
is small and thus the growth rate of the tilt will be small.
In addition, in the case of dwarf novae, the inclination of the orbital plane to the disk plane is small,
i.e., $\delta\sim 0$.
Hence, the tilts shown in the last three lines of table 1 are also not expected with observable amplitude.
In summary, the tilts which will be important in dwarf novae will be 
those of the second line ($\Omega_{\rm L}/\Omega_{\rm orb}=3$ and $r_{\rm L}/a=0.48$) and the seventh line
($\Omega_{\rm L}/\Omega_{\rm orb}=11.19$ and $r_{\rm L}/a=0.20$), and supplementally those of the third line
($\Omega_{\rm L}/\Omega_{\rm orb}=4$ and $r_{\rm L}/a=0.40$)
and the sixth line ($\Omega_{\rm L}/\Omega_{\rm orb}=7.46$ and $r_{\rm L}/a=0.26$).
Excitation of tilt at $\Omega_{\rm L}/\Omega_{\rm orb}=3$ is nothing but what was pointed out by Lubow (1992).
Numerical evaluation of growth rate [given by equation (\ref{growth})] in the above four cases will be made in future.

In Be/X-ray systems, the Be-star disk and the orbital plane of a compact star (neutron star in general)
is generally misaligned, and further a secondary star has a large excentric orbit.
In colliding or merging black-hole binaries, each black hole will have its own disk, 
although the system as a whole will be surrounded by a common envelope.
In Be/X-ray stars, for example, in addition to normal outbursts, giant outbursts which occur less
frequently are observed.
The giant outbursts are supposed to be due to an interaction between warped precessing Be-star disk and
secondary star which has a large eccentric orbit (Moritani et al. 2013).
Considering such situations, it will be important to examine the wave-wave resonant process in disks whose plane
is inclined from the orbital plane of an eccentric binary system.
In such systems resonant excitation of tilt at various radii listed in table 1 will have comparable 
growth rates.

\section{Discussions}

We have demonstrated that on tidally deformed disks, slowly retrograding tilts are generated by
the wave-wave resonant processes at various phases of disk temperature (see figure 4).
Our analyses in this paper, however, are only for special cases of $r_{\rm L1}=r_{\rm L2}$.
That is, we have assumed that for a trapped tilt mode whose capture radius is $r_{\rm L1}$,
there is always trapped $\omega_2$-oscillation whose capture radius $r_{\rm L2}$ is sufficiently close to 
$r_{\rm L1}$.
In real situations, there may  be no such $\omega_2$-oscillations, depending mainly on temperature 
distribution on disks.
Even in such cases, unless the difference between $r_{\rm L1}$ and $r_{\rm L2}$ is too large, the resonant
instability occurs,  although the growth rate will decrease with increase of the separation between
$r_{\rm L1}$ and $r_{\rm L2}$.
This comes from the following situations.
We can calculate the coupling factor, $W$ [see equation (\ref{2.1a})], of the resonant interaction and thus the
grwoth rate given by equation (\ref{growth}), 
if we know the eigenfunctions of $\omega_1$- and $\omega_2$-oscillations and the functional form of tidal waves,
even if $r_{\rm L1}\not= r_{\rm L2}$.
From conjucture in the case of zero teperature disks, however, we suppose that the growth rate is high in the
case of $r_{\rm L1}\sim r_{\rm L2}$.
Conversely, there will be cases where $r_{\rm L1}=r_{\rm L2}$ is realized for $\omega_1$ and $\omega_2$ which
are slightly deviated from the resonant condition, $\omega_1+\omega_2+\omega_{\rm D}=0$.
Even in such cases the resonant instability occurs, since in finite temperature disks, the resonant
instability is not sharply restricted only to the case of $\omega_1+\omega_2+\omega_{\rm D}=0$,
i.e., the resonance is broadened in frequency space.
Considering the above situations, we think that in real disks, the resonant instability occurs 
in finite regions around each discrete point on figure 4.

Next, it is noted why in the case of excitation of the tilt mode, we have much variety in the resonant radius 
and thus in the precession frequency of the tilt, 
compared with the case of the excitation of one-armed p-mode oscillation.\footnote{
The excitation of one-armed oscillation is considered to be the cause of superhumps in
dwarf novae.
}
Let us consider, for simplicity, the case of coplanar systems ($\delta=0$).
The oscillations describing the tilt is asymmetric with respect to the equatorial plane.
Hence, the $\omega_2$-oscillation must also be asymmetric ($n_1=1$) with respect to the equatorial
plane, since the tidal force is plane-symmetric ($\delta=0$) with respect to the
equatorial plane in coplanar cases.
This means that $n_2$ of the $\omega_2$-oscillation must be odd (not $n_2=0$).
Hence, if we consider a vertical p-mode oscillation as the $\omega_2$-oscillation, the mode has
a frequency given by $\omega_2=(m_2\Omega-\sqrt{n_2}\Omega_\bot)_{\rm L2}$, $n_2$ being 1, 3, ...
This frequency is not an integer times $\Omega_{\rm L2}$, and can have various values by
variety of $m_2$ and $n_2$.

Distinct from the above case of tilt excitation, 
in the case of excitation of the one-armed p-mode oscillation, the $\omega_2$-oscillation contributing to the 
excitation needs to be plane-symmetric, in the case of coplanar systems.
Hence, typical modes of the $\omega_2$-oscillation are those with $n_2=0$.
This makes the variety of $\omega_2$-oscillations which can contribute to resonance 
less than that in the case of excitation of tilts.


Observations of dwarf novae show that they have various types of time variations.
One of well-studied time variations are (positive) superhumps, which appear in the superoutburst phase.
The origin of their time variations is the excitation of low-frequency one-armed progressive 
p-mode oscillations (Osaki 1985) associated with the so-called 3 : 1 tidal instability
by Lubow (1991).
In addition to the superhumps, negative superhumps are often observed in various phases 
(from quiscent to superoutburst phases) of the disk evolution of dwarf novae.
Recently, for example, Osaki \& Kato (2013a) show that the frequency of the negative superhump varies systematically
during a supercycle (a cycle from one superoutburst to the next).
The negative superhumps have been supposed to be due to disk tilt, but the origin of the tilt seems not
to be understood yet, although many possibilities are suggested, e.g.,
the  tidal instability by Lubow (1992), magnetic couplings between the disk and either the secondary
or the primary (Murray et al. 2002), stream-disk interaction with variable vertical component of stream
due to asymmetric irradiation of the secondary (Smak 2009), and dynamical lift by the gas stream 
(Montgomery \& Martin 2010).
Our wave-wave resonant instability process is an extension of Lubow's
tidal instability one (Lubow 1991, 1992). 
We think that further examination  along this line is important to judge whether excitation of tilt by
the tidal instability can be
regarded as one of possible origins of negative superhumps.

\bigskip
The author thanks the referee for valuable suggestions, which much improved the paper.

\bigskip\noindent
{\bf Appendix 1.  Trapping (Capture) Condition of Tilt Mode by the WKB Method}

We introduce the variable $h_1$, defined by $h_1=p_1/\rho_0$, where $p_1$ is the pressure 
variation associated with oscillations over the unperturbed pressure $p_0(r,z)$, 
and $\rho_0(r,z)$ is the unperturbed density.
Here, $r$ and $z$ are the $r$- and $z$-components of the cylindrical coordinates whose center is at
the center of the primary star.
The radial wavelength of oscillations is assumed to be moderately short.
Then, we have a wave equation in terms of $h_1$ as (e.g., Kato 2001)
\begin{equation}
   \frac{1}{\rho_0}\frac{\partial}{\partial z}\biggr(\rho_0\frac{\partial h_1}{\partial z}\biggr)
   +(\omega-m\Omega)^2\frac{\partial}{\partial r}\biggr[\frac{1}{(\omega-m\Omega)^2-\kappa^2}
       \frac{\partial h_1}{\partial r}\biggr]
       +\frac{(\omega-m\Omega)^2}{c_{\rm s}^2}h_1=0,
\label{A1}
\end{equation}
where $c_{\rm s}(r)$ is the acoustic speed in the disk, and $m=1$ in the present problem,  
since we are considering the tilt mode here.
 
Wave equations similar to equation (\ref{A1}) have been solved by the WKB method, approximately
separating $h_1(r,z)$ as $h_1(r,z)=g(\eta)f(r,\eta)$ (e.g., Silbergleit et al. 2001, Kato 2012). 
Here, $\eta$ is defined by $\eta=z/H$ and  $H(r)$ is the scale height of the disk half thickness.
After separating wave equation (\ref{A1}) into two ordinary differential equations describing
variations in the vertical and horizontal directions by use of $g$ and $f$, 
we first solve the equation describing the variation in the vertical direction.
The equation is the Hermite equation in the case of vertically isothermal disks.
Boundary conditions at $z/H=\pm\infty$ gives us a disccrete set of the separation constant, $n$,
which is $n=0, 1,2,..$.(e.g., Okazaki et al. 1997). 
Then, the equation describing the variations in the horizontal direction becomes
\begin{equation}
    c_{\rm s}^2\frac{d}{dr}\biggr[\frac{1}{(\omega-m\Omega)^2-\kappa^2}\frac{df}{dr}\biggr]
        +\frac{(\omega-m\Omega)^2-n\Omega_\bot}{(\omega-m\Omega)^2}f =0,
\label{A2}
\end{equation}
where $n$ is the separation constant mentioned above and $n=1$ in the case of the tilt (and also $m=1$).

By introducing $\tau(r)$ defined by 
\begin{equation}
    \tau(r)=\int_{r_{\rm i}}^r \biggr[[\omega-m\Omega(r')]^2-\kappa^2(r')\biggr]dr',
\label{A3}
\end{equation}
where $r_{\rm i}$ is the radius where an inner boundary condition is imposed, 
we can reduce equation (\ref{A2}) to
\begin{equation}
   \frac{d^2f}{dr^2}+Qf=0,
\label{A4}
\end{equation}
where 
\begin{equation}
    Q(\tau)=\frac{(\omega-m\Omega)^2-n\Omega_\bot^2}
                 {c_{\rm s}^2(\omega-m\Omega)^2[(\omega-m\Omega)^2-\kappa^2]}.
\label{A5}
\end{equation}

Equation (\ref{A4}) is solved by the standard WBK method with the conditions that
the outer trapping (capture) radius, $r_{\rm L1}$, is a turning point of  $Q$, and at the inner boundary
the radial velocity vanishes.  
Then, the trapping condition is given by equation (\ref{2.6}).

\bigskip\noindent
{\bf Appendix 2.  Modes of Tidal Wave}

The values of $n(\equiv \omega_{\rm D}/\Omega_{\rm orb})$ and $m_{\rm D}$ required for 
the tidal waves have been summarized in table 1.
The subject to be considered here is whether such tidal waves really exist in tide and whether eccentricity
of the orbit is required.

We consider the tidal perturbations induced at a position ${\rm P}(\mbox{\boldmath $r$})$
on the disk of the primary by a scondary star of mass $M_{\rm s}$.
When the point P is at a distance $R[=(r^2+z^2)^{1/2}]$ from the center of the primary and the secondary star's
zenith distance observed at the point P is $\vartheta$ (see figure 5), the tidal gravitational
potential $\psi_{\rm D}(\mbox{\boldmath $r$}, t)$  at the point P is given by
(e.g., Lamb 1924)
\begin{equation}
    \psi_{\rm D}=-\frac{GM_{\rm s}}{(D^2-2RD{\rm cos}\vartheta+R^2)^{1/2}}
       +\frac{GM_{\rm s}}D^2R\ {\rm cos}\vartheta,
\label{psiD1}
\end{equation}
where $D$ is the distance between the primary and secondary stars at time $t$.
The second term on the right-hand side represents the potential of a uniform field of force of the
secondary acting on the primary.
Since $D$ is larger than $R$, the right-hand side of equation (\ref{psiD1}) is expanded 
by a power series of $R/D$ as 
\begin{equation}
   \frac{\psi_{\rm D}}{GM_{\rm s}/D}=-1-\biggr(\frac{R}{D}\biggr)^2P_2({\rm cos}\vartheta)
             -\biggr(\frac{R}{D}\biggr)^3P_3({\rm cos}\vartheta)-...,
\label{psiD2}
\end{equation}
where $P_2({\rm cos}\vartheta)$ and $P_3({\rm cos}\vartheta)$ are the Legendre polynomials
$P_\ell$ of argument ${\rm cos}\vartheta$ with $\ell=2$ and $\ell=3$, respectively.
In the case where the orbit of the secondary is elliptical, $D$ is a function of time.
Hence, it is convenient to normalize $\psi_{\rm D}$ by a time-independent quantity.
Here, we normalize $\psi_{\rm D}$ by the mean radius, $a$, of the elliptical orbit
(see later for the relation between $D$ and $a$) as
\begin{equation}
   \frac{\psi_{\rm D}}{GM_{\rm s}/a}=-\frac{a}{D}
        -\biggr(\frac{R}{a}\biggr)^2\biggr(\frac{a}{D}\biggr)^3P_2({\rm cos}\vartheta)
        -\biggr(\frac{R}{a}\biggr)^3\biggr(\frac{a}{D}\biggr)^4P_3({\rm cos}\vartheta)-...,
\label{psiD3}
\end{equation}

\begin{figure}
\begin{center}
    \FigureFile(80mm,80mm){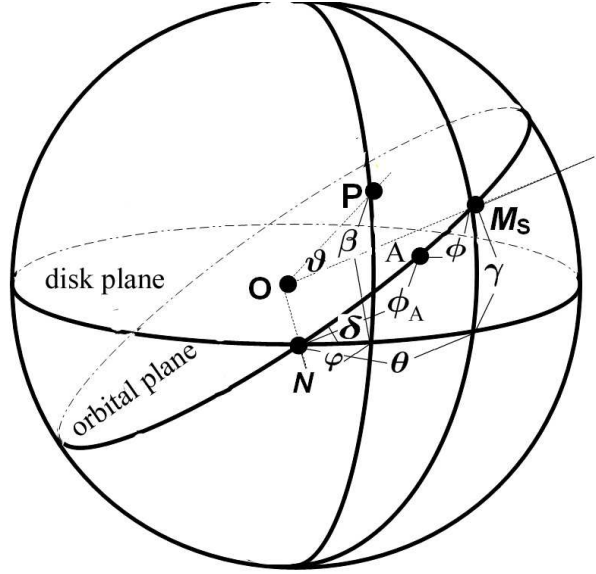}
\end{center}
\caption{Relation between disk plane and orbital plane of the secondary 
}
\end{figure}

The next problem is to represent $P_2({\rm cos}\vartheta)$ [and $P_3({\rm cos}\vartheta)$] 
in terms of the cylindrical coordinates ($r$, $\varphi$, $z$) of the point P and coordinates ($\theta$, $\gamma$)
representing the position of the secondary on the revolutional orbit.
As a preparation, we consider a unit sphere whose center is the center of the primary,
as is shown in figure 5.
The poles of the sphere are taken in the direction perpendicular to the disk plane.
The orbital plane of the secondary inclines to the disk plane by angle $\delta$.
Let us denote the spherical coordinates of the point P by $\varphi$ and $\beta$, as shown in figure 5.
The angle $\varphi$ is measured from the nodal point $N$.
Then, using a formula of spherical trigonometry we have
\begin{equation}
   {\rm cos}\vartheta={\rm sin}\beta\ {\rm sin}\gamma+{\rm cos}\beta\ {\rm cos}\gamma\ {\cos}(\theta-\varphi).
\label{formula1}
\end{equation}
Hence, simple algebraic calculations give
\begin{eqnarray}
     P_2({\rm cos}\vartheta)=&&\frac{1}{2}(3\ {\rm cos}^2\vartheta-1)  \nonumber \\
         =&&\frac{1}{4}(3\ {\rm sin}^2\beta-1)(3\ {\rm sin}^2\gamma-1)+\frac{3}{4}{\rm sin}2\beta\ {\rm sin}2
                     \gamma\ {\rm cos}(\theta-\varphi)  \nonumber \\
         && +\frac{3}{4}{\rm cos}^2\beta\ {\rm cos}^2\gamma\ {\rm cos}[2(\theta-\varphi)],
\label{P2}
\end{eqnarray}
and
\begin{eqnarray}
      P_3({\rm cos}\vartheta)=&&\frac{1}{2}(5\ {\rm cos}^2\vartheta-3\ {\rm cos}\vartheta)   \nonumber \\
       =&&\frac{1}{4}\ {\rm sin}\beta\ {\sin}\gamma\biggr[25\ {\rm sin}^2\beta\ {\rm sin}^2\gamma-15({\rm sin}^2\beta+{\rm sin}^2\gamma)+9\biggr]  \nonumber \\
        &&+\frac{1}{8}\ {\rm cos}\beta\ {\rm cos}\gamma\biggr[75\ {\rm sin}^2\beta\ {\rm sin}^2\gamma+3-15({\rm sin}^2\beta+{\rm sin}^2\gamma)\biggr]{\rm cos}(\theta-\varphi)  \nonumber  \\
        &&+\frac{15}{4}\ {\rm sin}\beta\ {\rm sin}\gamma\ {\rm cos}^2\beta\ {\rm cos}^2\gamma\ {\rm cos}[2(\theta-\varphi)]   \nonumber   \\
        &&+\frac{5}{8}\ {\rm cos}^3\beta\ {\rm cos}^3\gamma\ {\rm cos}[3(\theta-\varphi)],
\label{P3}
\end{eqnarray}
where  
$P_2({\rm cos}\vartheta)$ and $P_3({\rm cos}\vartheta)$ have been expressed in terms of 
$\beta$, $\gamma$, and $\theta-\varphi$.

Here, $\theta$ and $\gamma$ are related to $\delta$ and $\phi_{\rm A}+\phi$ by spherical trigonometric 
formulae:
\begin{equation}
     {\rm cos}\gamma={\rm cos}(\phi_{\rm A}+\phi)\ {\rm cos}\theta+{\rm sin}(\phi_{\rm A}+\phi)\ {\rm sin}\theta\ {\rm cos}\delta,
    \quad     {\rm sin}\gamma={\rm sin}(\phi_{\rm A}+\phi)\ {\rm sin}\delta,
\label{formula2}
\end{equation}
where $\phi_{\rm A}$ is the angular direction of periastron, $A$, measured from the nodal point $N$ along
the orbit of the secondary, and $\phi$ is the position of the secondary on the orbit, 
measured from the periastron, $A$ (see figure 5).
These relations (\ref{formula2}) show that in the limit of $\delta =0$, we have $\gamma=0$ and
$(\phi_{\rm A}+\phi)=\theta$, as expected.
Even when $\delta\not= 0$, the above relations (\ref{formula2}) show that ${\rm sin}\gamma
=\delta\ {\rm sin}(\phi_{\rm A}+\phi)$ and $\phi_{\rm A}+\phi=\theta$ 
until the order of $\delta^2$.
Hence, in this paper, assuming that the misalignement between the disk and orbital planes is not large,
we adopt
\begin{equation}
            \theta=(\phi_{\rm A}+\phi), \quad {\rm sin}\gamma=\delta\ {\rm sin}(\phi_{\rm A}+\phi),
\label{theta}
\end{equation}
and the terms of the order of $\delta^2$ are neglected in equations (\ref{P2}) and (\ref{P3}).
Then, $P_2({\rm cos}\ \vartheta)$ and $P_3({\rm cos}\ \vartheta)$ are approximated as 
\begin{eqnarray}
    P_2({\rm cos}\vartheta)\sim&&\frac{3}{4}(1-3\ {\rm sin}^2\beta)
          +\frac{3}{2}\delta\ {\rm sin}2\beta\biggr[{\rm sin}(2\phi+2\phi_{\rm A}-\varphi)+{\rm sin}\varphi\biggr]
           \nonumber \\
      +&&\frac{3}{4}{\rm cos}^2\beta\ {\rm cos}\biggr[2(\phi+\phi_{\rm A}-\varphi)\biggr],
\label{P_2approx}
\end{eqnarray}
and
\begin{eqnarray}
     P_3({\rm cos}\vartheta)\sim&& -\frac{3}{4}\delta\ {\rm sin}\ \beta\ (5{\rm sin}^2\beta-3)
                    \ {\rm sin}(\phi+\phi_{\rm A})
         +\frac{1}{8}(3-15\ {\rm sin}^2\beta)\ {\rm cos}\beta\ {\rm cos}(\phi+\phi_{\rm A}-\varphi)
                 \nonumber   \\
         +&&\frac{15}{8}\delta\ {\rm sin}\beta\ {\rm cos}^2\beta
           \biggr[{\rm sin}(3\phi+3\phi_{\rm A}-2\varphi)-{\rm sin}(\phi+\phi_{\rm A}-2\varphi)\biggr]
                 \nonumber  \\
         +&&\frac{5}{8}{\rm cos}^3\beta\ {\rm cos}\biggr[3(\phi+\phi_{\rm A}-\varphi)\biggr].
\label{P_3approx}
\end{eqnarray}

In the limiting case  where the secondary star's orbit is circular (i.e., $e=0$), 
$\phi$ is obviously $\phi=\Omega_{\rm orb}t$.
Hence, in this case, if the orbital plane coincides with the disk plane (i.e., $\delta=0$), 
the tidal waves are two-armed ($m_{\rm D}=2$) with frequency $2\Omega_{\rm orb}$ (i.e., $\omega_{\rm D}=
2\Omega_{\rm orb}$) [see equation (\ref{P_2approx}) and table 2],  
if the expansion in equation (\ref{psiD2}) is terminated by the second term on the right-hand side.
If the expansion proceeds till the next term, 
the one-armed ($m_{\rm D}=1$) tidal waves with frequency $\Omega_{\rm orb}$, and 
the three-armed ($m_{\rm D}=3$) tidal waves with frequency $3\Omega_{\rm orb}$ 
appear [see equation (\ref{P_3approx}) and table 2],  
but the ratio of $\omega_{\rm orb}/m_{\rm D}$ is still $\Omega_{\rm orb}$.
In misaligned cases ($\delta\not= 0$), the situations are changed and wven when $e=0$,  
we have one-armed ($m_{\rm D}=1$) tidal waves with $\omega_{\rm orb}=2\Omega_{\rm orb}$ 
[see equation (\ref{P_2approx}) and table 3], and 
two-armed ($m_{\rm D}=2$) waves with $\omega_{\rm D}=\Omega_{\rm orb}$ and $\omega_{\rm D}=3\Omega_{\rm orb}$
[see equation (\ref{P_3approx}) and table 3].
That is, in these cases the ratio, $\omega_{\rm D}/m_{\rm D}$, is not always $\Omega_{\rm orb}$, but
the ratio of $2\Omega_{\rm orb}$ and $(3/2)\Omega_{\rm orb}$ appears even if $e=0$. 

Some tidal waves required to lead to resonant instability (see table 1), however, are still different 
from those tidal waves mentioned above.
That is, the results in section 5 show that for some resonant instability to occur, other relations  
of $\omega_{\rm D}/m_{\rm D}$ are required.
For such tidal waves to occur, the orbit of the secondary is needed to be eccentric.
To examine this in detail, we must know i) the deviation of $\phi(t)$ from $\phi=\Omega_{\rm orb}t$ 
and ii) the time variation of $a/D$ in the cases of elliptical orbits.

The functional form of $\phi(t)$ is well known in the fields of celestial mechanics.
The main results concerning $\phi(t)$ are summarized as follows. 
Let the eccentricity and the mean radius of the orbit be $e$ and $a$, respectively.
Then, the distance of the secondary from the center of the primary, $D$, changes with a change of 
$\phi(t)$
as
\begin{equation}
      D=\frac{a(1-e^2)}{1+e\ {\rm cos}\phi}.
\label{D}
\end{equation}
Now, we introduce an angle $u$ (eccentric anomaly) defined by 
\begin{equation}
      D=a(1-e\ {\rm cos} u).
\label{D-eq}
\end{equation}
Then, the equation of motion shows that the time variation of $u(t)$ is described by 
the Kepler equation:
\begin{equation}
    u-e\ {\rm sin} u=\Omega_{\rm orb}t,
\label{u-eq}
\end{equation}
where $u=0$ (and thus $\phi=0$) is taken at $t=0$.
In the limit of the circular orbit ($e=0$), we have $D=a$ and $u=\phi=\Omega_{\rm orb}t$.

Equation (\ref{u-eq}) is solved with respect to $u$ by a power series of $e$, assuming that $e$ is small.
Then, we have (e.g., Araki 1980)
\begin{eqnarray}
     u=&&\Omega_{\rm orb}t+e\ {\rm sin}(\Omega_{\rm orb}t)+\frac{1}{2}e^2{\rm sin}(2\Omega_{\rm orb}t)
                                 \nonumber   \\
       &&+\frac{1}{8}e^3\biggr[3\ {\rm sin}(3\Omega_{\rm orb}t)-{\rm sin}(\Omega_{\rm orb}t)\biggr] +...
\label{u-eq2}
\end{eqnarray}
Combination of equations (\ref{D}) and (\ref{D-eq}) gives $(1+e\ {\rm cos}\ \phi)(1+e\ {\rm cos}\ u)
=(1-e^2)$.
From this equation and $\phi=\Omega_{\rm orb}t$ in the limit of $e=0$, we obtain 
(e.g., Araki 1980)
\begin{eqnarray}
    \phi=&&\Omega_{\rm orb}t+2e\ {\rm sin}(\Omega_{\rm orb}t)
              +\frac{5}{4}e^2{\rm sin}(2\Omega_{\rm orb}t)    \nonumber  \\
         &&+\frac{1}{12}e^3\biggr[13{\rm sin}(3\Omega_{\rm orb}t)-3{\rm sin}(\Omega_{\rm orb}t)\biggr]+...
\label{tildphi}
\end{eqnarray}
Furthermore, $a/D=(1-e\ {\rm cos}\ u)^{-1}$ is also expanded by a power series of $e$ as
\begin{equation}
   \frac{a}{D}=1+e\ {\rm cos} u+e^2\ {\rm cos}^2 u+e^3\ {\rm cos}^3 u+...
\label {aoverD}
\end{equation}
Then, by using equation (\ref{u-eq2}) we can write $a/D$ explicitly as a function of $\Omega_{\rm orb}t$:
\begin{eqnarray}
     \frac{a}{D}=&&1+e\ {\rm cos}(\Omega_{\rm orb}t)+e^2{\rm cos}(2\Omega_{\rm orb}t)
                         \nonumber   \\
                 &&+\frac{1}{8}e^3\biggr[9\ {\rm cos}(3\Omega_{\rm orb}t)-{\rm cos}(\Omega_{\rm orb}t)\biggr].
\label{aoverD-2}
\end{eqnarray}

If expressions for $P_\ell$'s [equations (\ref{P_2approx}) and (\ref{P_3approx})], $a/D$ [equation (\ref{aoverD-2})],
and $\phi$ [equation (\ref{tildphi})] are substituted into equation $\psi_{\rm D}$ [equation (\ref{psiD3})],
we  have $\psi_{\rm D}$ expressed directly in terms of the coordinates of the observing point, ($\varphi$, $\beta$),
the secondary's orbital parameters ($\Omega_{\rm orb}$, $e$, $a$) and the inclination, $\delta$,
between the disk and orbital planes.
The tidal waves have generally  forms of ${\rm cos}\ [n\Omega_{\rm orb}-m_{\rm D}\varphi]$ or 
${\rm sin}\ [n\Omega_{\rm orb}-m_{\rm D}\varphi]$.
The phase $n\Omega_{\rm orb}-m_{\rm D}\varphi$ in various cases is shown in tables 2 and 3,
by taking $m_{\rm D}>0$.

\begin{table}
  \caption{Time and Azimuthal Dependences of Tidal Waves in Coplanar Systems.}\label{tab:second}
  \begin{center}
    \begin{tabular}{ccrr}
      \hline
    Separation & $e^0$  & $e^1$ & $e^2$ \\
        \hline\hline
    $(R/a)^2$ &  $2\Omega_{\rm orb}t-2\varphi$ &  $\Omega_{\rm orb}t-2\varphi$  &                   $-2\varphi$ \\
              &                                & $3\Omega_{\rm orb}t-2\varphi$  & $2\Omega_{\rm orb}t-2\varphi$ \\
              &                                &                                & $4\Omega_{\rm orb}t-2\varphi$ \\
  \hline
    $(R/a)^3$ &  $\Omega_{\rm orb}t-\varphi$   &                   $-\varphi$   & $-\Omega_{\rm orb}t-\varphi$  \\
              &                                & $2\Omega_{\rm orb}t-\varphi$   & $\Omega_{\rm orb}t-\varphi$   \\
              &                                &                                & $3\Omega_{\rm orb}t-\varphi$  \\
              &  $3\Omega_{\rm orb}t-3\varphi$ & $2\Omega_{\rm orb}t-3\varphi$  & $\Omega_{\rm orb}t-3\varphi$  \\
              &                                & $4\Omega_{\rm orb}t-3\varphi$  & $3\Omega_{\rm orb}t-3\varphi$ \\
              &                                &                                & $5\Omega_{\rm orb}t-3\varphi$ \\
  \hline
    $(R/a)^4$ &  $2\Omega_{\rm orb}t-2\varphi$ & $\Omega_{\rm orb}t-2\varphi$   &                   $-2\varphi$ \\
              &                                & $3\Omega_{\rm orb}t-2\varphi$  & $2\Omega_{\rm orb}t-2\varphi$ \\
              &                                &                                & $4\Omega_{\rm orb}t-2\varphi$ \\
              &  $4\Omega_{\rm orb}t-4\varphi$ & $3\Omega_{\rm orb}t-4\varphi$  & $2\Omega_{\rm orb}t-4\varphi$ \\
              &                                & $5\Omega_{\rm orb}t-4\varphi$  & $4\Omega_{\rm orb}t-4\varphi$ \\
              &                                &                                & $6\Omega_{\rm orb}t-4\varphi$ \\
\hline
    \end{tabular}
  \end{center}
\end{table}%

\begin{table}
  \caption{Time and Azimuthal Dependences of Tidal Waves Added in Coplanar Systems with $\delta\ll 1$.}\label{tab:third}
  \begin{center}
    \begin{tabular}{crrr}
      \hline
    Separation & $e^0$  & $e^1$ & $e^2$ \\
        \hline\hline
    $(R/a)^2$ &                    $-\varphi$  & $-\Omega_{\rm orb}t-\varphi$   & $-2\Omega_{\rm orb}t-\varphi$\\
              &                                & $ \Omega_{\rm orb}t-\varphi$   &                    $-\varphi$ \\
              &                                &                                & $2\Omega_{\rm orb}t-\varphi$ \\
              &  $2\Omega_{\rm orb}t-\varphi$  &  $\Omega_{\rm orb}t-\varphi$   &                   $-\varphi$  \\
              &                                & $3\Omega_{\rm orb}t-\varphi$   & $2\Omega_{\rm orb}t-\varphi$ \\
              &                                &                                & $4\Omega_{\rm orb}t-\varphi$ \\  \hline
    $(R/a)^3$ &  $\Omega_{\rm orb}t-2\varphi$  &                   $-2\varphi$  & $-\Omega_{\rm orb}t-2\varphi$ \\
              &                                & $2\Omega_{\rm orb}t-2\varphi$  & $\Omega_{\rm orb}t-2\varphi$   \\
              &                                &                                & $3\Omega_{\rm orb}t-2\varphi$  \\
              &  $3\Omega_{\rm orb}t-2\varphi$ & $2\Omega_{\rm orb}t-2\varphi$  & $\Omega_{\rm orb}t-2\varphi$  \\
              &                                & $4\Omega_{\rm orb}t-2\varphi$  & $3\Omega_{\rm orb}t-2\varphi$ \\
              &                                &                                & $5\Omega_{\rm orb}t-2\varphi$ \\
  \hline
    $(R/a)^4$ &  $2\Omega_{\rm orb}t-\varphi$ & $\Omega_{\rm orb}t-\varphi$     &                   $-\varphi$ \\
              &                                & $3\Omega_{\rm orb}t-\varphi$   & $2\Omega_{\rm orb}t-\varphi$ \\
              &                                &                                & $4\Omega_{\rm orb}t-2\varphi$ \\
              &  $2\Omega_{\rm orb}t-3\varphi$ & $\Omega_{\rm orb}t-3\varphi$   &                   $-3\varphi$ \\
              &                                & $3\Omega_{\rm orb}t-3\varphi$  & $2\Omega_{\rm orb}t-3\varphi$ \\
              &                                &                                & $4\Omega_{\rm orb}t-3\varphi$ \\
              &  $4\Omega_{\rm orb}t-3\varphi$ & $3\Omega_{\rm orb}t-3\varphi$  & $2\Omega_{\rm orb}t-3\varphi$ \\
              &                                & $5\Omega_{\rm orb}t-3\varphi$  & $4\Omega_{\rm orb}t-3\varphi$ \\
              &                                &                                & $6\Omega_{\rm orb}t-3\varphi$ \\
  \hline
    \end{tabular}
  \end{center}
\end{table}

\bigskip
\leftskip=20pt
\parindent=-20pt
\par
{\bf References}
\par
Araki, T. 1980, M\'{e}canique C\'{e}leste (Japanese) (Kooseisha, Tokyo), pp. 85-86    \par
Ferreira, B., \& Ogilvie, G.I. 2008, MNRAS, 386, 2297\par
Kato, S. 2001, PASJ, 53, 1\par 
Kato, S., 2004, PASJ, 56, 905\par
Kato, S. 2008, PASJ, 60, 111 \par
Kato, S. 2012, PASJ, 64, 78 \par
Kato, S. 2013a, PASJ, 65, 56 \par
Kato, S. 2013b, PASJ, 65, 75 \par
Kato, S. 2013c, submitted to PASJ\par
Kato, S., Fukue, J., \& Mineshige, S. 2008, Black-Hole Accretion Disks --- Towards a New paradigm --- 
  (Kyoto: Kyoto University Press), chap. 11 \par
Kato, S., Okazaki, A.T.,\& Oktariani, F. 2011, 63, 363 \par
Lamb, H. 1924, Hydrodynamics (Cambdridge University Press, Cambridge), p.336 \par
Lubow, S. H. 1991, ApJ, 381, 259 \par
Lubow, S. H. 1992, ApJ, 401, 317\par
Meyer-Vernet, N. \& Sicardy, B. 1987, Icarus, 69, 157\par
Montgomery, M.M., \& Martin, E.L. 2010, ApJ, 722, 989\par
Moritani, Y., Nogami, D., Okazaki, A.T., Imada, A., Kambe, E., Honda, S., Hashimoto, O., Mizoguchi, 
    S., Kanda, Y., Sadakane, K., Ichikawa, K.,
    2013, PASJ, 65,83\par
Murray, J.R., Chakrabarty, D., Wynn, G.A., \& Kramer, L. 2002, MNRAS, 335, 247 \par 
Oktarian, F., Okazaki, A.T., Kato, S. 2010, PASJ, 62, 709 \par
Ohshima, T., et al. 2012, PASJ, 64, L3 \par
Okazaki, A.T., Kato, S., \& Fukue, J. 1987, PASJ, 39, 457\par
Osaki, Y. 1985, A\&A, 144, 369 \par 
Osaki, Y., Kato, T. 2013a, PASJ, 65, 50\par
Osaki, Y., Kato, T. 2013b, PASJ, 65, in press \par 
Patterson, J., Jablonski, F., Koen, C., O'Donoghue, D., \& Skillman, D.R. 1995, PASP, 107,1183\par
Silbergleit, A.S., Wagoner, R.V., Ortega-Rodr\'{i}gues, M. 2001, ApJ, 548, 385\par
Smak, J. 2009, Acta Astron., 59, 419\par

\leftskip=0pt
\parindent=0pt
\bigskip\noindent

\end{document}